\pdfoutput=1
\RequirePackage{ifpdf}
\ifpdf 
\documentclass[pdftex]{sigma}
\else
\documentclass{sigma}
\fi

\usepackage{bbold}

\begin{document}

\allowdisplaybreaks

\renewcommand{\thefootnote}{$\star$}

\renewcommand{\PaperNumber}{082}

\FirstPageHeading

\ShortArticleName{Solutions of the Dirac Equation in a Magnetic Field and Intertwining Operators}

\ArticleName{Solutions of the Dirac Equation in a Magnetic Field\\ and Intertwining Operators\footnote{This
paper is a contribution to the Special Issue ``Superintegrability, Exact Solvability, and Special Functions''. The full collection is available at \href{http://www.emis.de/journals/SIGMA/SESSF2012.html}{http://www.emis.de/journals/SIGMA/SESSF2012.html}}}

\Author{Alonso CONTRERAS-ASTORGA~$^\dag$, David J. FERN\'ANDEZ C.~$^\dag$ and Javier NEGRO~$^\ddag$}

\AuthorNameForHeading{A.~Contreras-Astorga, D.J.~Fern\'andez~C. and J.~Negro}

\Address{$^\dag$~Departamento de F\'isica, Cinvestav, AP 14-740, 07000 M\'exico DF, Mexico}
\EmailD{\href{mailto:acontreras@fis.cinvestav.mx}{acontreras@fis.cinvestav.mx}, \href{mailto:david@fis.cinvestav.mx}{david@fis.cinvestav.mx}}

\Address{$^\ddag$~Departamento de F\'isica Te\'orica, At\'omica y \'Optica, Universidad de Valladolid,\\
\hphantom{$^\ddag$}~47071 Valladolid, Spain}
\EmailD{\href{mailto:jnegro@fta.uva.es}{jnegro@fta.uva.es}}

\ArticleDates{Received July 31, 2012, in f\/inal form October 17, 2012; Published online October 28, 2012}

\Abstract{The intertwining technique has been widely used to study the Schr\"odinger equation and to generate new Hamiltonians with known
spectra. This technique can be adapted to f\/ind the bound states of certain Dirac Hamiltonians. In this paper the system to be solved is a relativistic particle placed in a magnetic f\/ield with cylindrical symmetry whose intensity decreases as the distance to the symmetry axis grows and its f\/ield lines are parallel to the $x-y$ plane. It will be shown that the Hamiltonian under study turns out to be shape invariant.}


\Classification{81Q05; 81Q60; 81Q80}

\renewcommand{\thefootnote}{\arabic{footnote}}
\setcounter{footnote}{0}

\vspace{-2mm}

\section{Introduction}

The intertwining technique, also called Supersymmetric Quantum Mechanics (SUSY QM), is a~widespread method used to generate exactly solvable Hamiltonians departing from a given initial one and can be employed as well to solve a certain set of Hamiltonians in a closed way, among other applications. In the simplest case (1-SUSY QM) the new potentials have similar spectra as the original one, namely, they might dif\/fer at most in the ground state energy. Examples of potentials generated by this technique are those which arise when adding a bound state to the free particle Hamiltonian (hyperbolic P\"oschl--Teller)~\cite{Sukumar04} or the Abraham--Moses--Mielnik potentials which are isospectral to the harmonic oscillator \cite{Samsonov97, Junker98, Bogdan84}. This method has been also applied successfully to the radial part of the hydrogen atom potential \cite{Samsonov97,Fernandez84, Junker98, Rosas98}, the trigonometric P\"oschl--Teller potentials \cite{Contreras08}, among many others.

To apply the technique \cite{Fernandez05} we start from two one-dimensional Schr\"odinger Hamiltonians
\begin{gather*}
H_i=-{1 \over 2}{\mathrm{d}^2 \over \mathrm{d}x^2}+V_i(x), \qquad i=0,1,
\end{gather*}
where $H_0$ is known. Now let us suppose the existence of a dif\/ferential operator $A^\dag_1$ which satisf\/ies
\begin{gather}
H_1 A^\dag_1 = A^\dag_1 H_0, \qquad A^\dag_1={1\over \sqrt{2}} \left( -{\mathrm{d} \over \mathrm{d}x}+W_1(x)\right). \label{entrelazamiento}
\end{gather}
Since the operator $A^\dag_1$ is of f\/irst order, the technique is known as 1-SUSY QM and the function $W_1(x)$ as the superpotential. It is also said that the potentials $V_0(x)$ and~$V_1(x)$, whose Hamiltonians are intertwined by the operator~$A^\dag_1$, are supersymmetric partners.

In order to satisfy equation~(\ref{entrelazamiento}) $V_1(x)$ and $W_1(x)$ must obey
\begin{gather}
V_1(x)= V_0(x)-W'_1(x), \qquad W_1'(x)+W_1^2(x)=2(V_0-\epsilon_1), \label{ricati}
\end{gather}
where $\epsilon_1$ is a real integration constant called factorization energy. From the previous equations we can see that if $V_0(x)$ is given and $W_1(x)$ is found, the supersymmetric partner $V_1(x)$ is completely determined. Furthermore, equation~(\ref{entrelazamiento}) ensures that if $\psi_n$ is an eigenfunction of $H_0$ with eigenvalue $E_n$ then $A_1^\dag \psi_n$ will be an eigenfunction of $H_1$ with the same eigenvalue. Note that the operators $A_1^\dag $ and $ (A_1^\dag)^\dag\equiv A_1 $ factorize the Hamiltonian as follows
\begin{gather}
H_0 = A_1 A_1^\dag +\epsilon_1, \qquad H_1 = A_1^\dag A_1+\epsilon_1, \label{factorizacion}
\end{gather}
where
\begin{gather*}
A_1={1\over \sqrt{2}} \left( {\mathrm{d} \over \mathrm{d}x}+W_1(x)\right).
\end{gather*}

By taking the squared norm of the vectors $A_1^\dag \psi_n$ we have $ | | A_1^\dag \psi_n | |^2 = \langle A_1^\dag \psi_n, A_1^\dag \psi_n \rangle = \langle \psi_n, A_1 A_1^\dag \psi_n \rangle = E_n - \epsilon_1 \geq 0 $ $\forall\, n$ which implies that $ \epsilon_1 \leq E_0$, where $E_0$ is the ground state energy of $H_0$. One could ask now if $\{A_1^\dag \psi_n, \;  n = 0,1,2, \dots \}$ is a complete orthogonal set. In order to answer this question, assume the existence of a vector $\psi_{\epsilon_1}$
orthogonal to each vector of the previous set, then
\begin{gather*}
\langle \psi_{\epsilon_1} , A_1^\dag\psi_n \rangle = \langle A_1 \psi_{\epsilon_1}, \psi_n \rangle = 0 \quad  \forall\, n \qquad \Rightarrow \qquad A_1 \psi_{\epsilon_1} = 0,
\end{gather*}
since $\left\{ \psi_n, \; n= 0,1,2,\dots \right\}$ is a complete orthogonal set. The f\/irst-order dif\/ferential equation $A_1 \psi_{\epsilon_1} = 0$
can be solved immediately
\begin{gather*}
\psi_{\epsilon_1} \propto \exp\left[-\int_0^x W_1(y) \mathrm{d}y \right].
\end{gather*}
Notice that $\psi_{\epsilon_1}$ satisf\/ies
\begin{gather*}
H_1 \psi_{\epsilon_1} = \epsilon_1 \psi_{\epsilon_1} .
\end{gather*}
Thus, depending on the square integrability of this vector, and the value of $\epsilon_1$, three possibilities arise
\begin{itemize}\itemsep=0pt

\item The function $\psi_{\epsilon_1}$ with $\epsilon_1 < E_0$ belongs to the Hilbert space $\mathcal{H}$. Thus, $\{ \psi_{\epsilon_1}, A^\dag_1 \psi_n, \; n=0,1,2,\dots \}$ is a complete orthogonal set, and from equations~(\ref{entrelazamiento}), (\ref{factorizacion}) the spectrum of $H_1$ is given by Sp$[H_1]= \left\{ \epsilon_1, E_n, \; n=0,1,2,\dots \right\}$.

\item $\psi_{\epsilon_1} \notin \mathcal{H}$ with $\epsilon_1 < E_1$. In this case $\{ A^\dag_1 \psi_n, \; n=0,1,2,\dots \}$ is a complete orthogonal set and thus Sp$[H_1]= \mathrm{Sp}[H_0]$.

\item When $\psi_{\epsilon_1} \notin \mathcal{H}, \epsilon_1 = E_0$, the set  $\{ A^\dag_1 \psi_n, \; n=1,2,3,\dots\}$ is complete and thus Sp$[H_1]= \{ E_n, \; n=1,2,3, \dots \}$.
\end{itemize}

Restricting ourselves to this last case, it can be verif\/ied that $W_1(x)=\psi_0' / \psi_0$ fulf\/ills equation~(\ref{ricati}) and applying successively this technique we can generate a hierarchy of Hamiltonians, where $ \mathrm {Sp}[H_0] \supset \mathrm{Sp}[H_1] \supset \mathrm{Sp}[H_2]\supset\cdots $. This sequence is either f\/inite or inf\/inite if the number of bound states of $ H_0 $ is f\/inite or inf\/inite respectively~\cite{Sukumar85}.

Up to this point we have assumed that starting from a solvable Hamiltonian $H_0$ we can ge\-ne\-rate a hierarchy of Hamiltonians $\left\{ H_i, \; i=0,1, \dots \right\}$. However, from the equation adjoint to equation~(\ref{entrelazamiento}) we can see that beginning from an eigenvector of $H_i $ we can construct an eigenvector of $H_{i-1}$ through the action of the operator $A_i$. Thus, if we had known enough eigenvectors of the Hamiltonians of the hierarchy, for example all the ground states, it would be possible to build all bound states of $H_0$ by applying the operators~$A_i$ over them.

It is convenient to recall now the concept of shape invariance.  {\it If two SUSY partner potentials $V_{1,2}(x;a_1)$ satisfy the condition
\begin{gather*}
V_2(x;a_1)=V_1(x;a_2)+R(a_1),
\end{gather*}
where $a_1$ is a set of parameters, $a_2$ is a function of $a_1$ and the remainder $R(a_1)$ is independent of $x$, then $V_1(x;a_1)$ and $V_2(x;a_1)$ are said to be shape invariant} \cite{Khare95}.

If the potentials of a hierarchy of Hamiltonians are shape invariant and the ground state of one of them is found, in principle, all the ground states can be derived. In this way we can f\/ind the eigenfunctions of the f\/irst Hamiltonian. The harmonic oscillator and the radial ef\/fective potential of the hydrogen atom are examples of shape invariant potentials that can be solved through this procedure.

It is noteworthy that there are papers in which through the SUSY technique the Dirac equation for dif\/ferent systems has been solved \cite{castaños91,Lima04,Samsonov02,Plyushchay12, Samsonov03, Axel10, Sukumar85b} or analyzed~\cite{Plyushchay11}. However, the dif\/ferences with respect to the approach we will use here will be signif\/icant.

In Section~\ref{sec:NRQM} we will employ the 1-SUSY QM in order to solve the stationary Schr\"odinger equation for a charged particle placed in a magnetic f\/ield generated by the vector potential $\vec{A}(\rho,\phi,z) = {c k \over e \rho} \hat{e}_z$, where $k$ is a constant characterizing the f\/ield strength, $c$ is the speed of light, $e$ is the charge of the particle, and $\rho$ is the radial variable in cylindrical coordinates. The resulting f\/ield has cylindrical symmetry, its intensity decreases with the distance to the symmetry axis and its f\/ield lines are parallel to the $x-y$ plane. Making use of the basic ideas to solve the shape invariant potentials through the 1-SUSY QM technique, we will work out the same problem in Section~\ref{sec:RQM} in the relativistic regime by solving the associated Dirac equation. In the last section we will present our conclusions.

\section{Nonrelativistic quantum approach}\label{sec:NRQM}

The classical Hamiltonian of a particle with charge $e$ and mass $m$ in a magnetic f\/ield generated by the vector potential  $\vec{A}(\rho,\phi,z) = {c k \over e \rho} \hat{e}_z$ is given by
\begin{gather*}
H_{\rm cl}={P_x^2 \over 2m}+{P_y^2 \over 2m} +{1 \over 2m}\left(P_z-{k \over \rho} \right)^2.
\end{gather*}
The corresponding magnetic f\/ield $\vec{B} = \nabla \times \vec{A}$ could
be produced in a coaxial transmission line, with the inner and outer
conductors carrying currents $I_a$ and $-I_a  \tau / (r_o+\tau)$ respectively, where $r_o$ is the minor radius of the outer conductor and $\tau$ is its thickness. If  the current density in the second conductor is $||\vec{J}|| = r_o I_a /(2 \pi \rho^3)$ then such a magnetic f\/ield will be
generated in the material~\cite{griffiths98, sadiku00}.

In order to address the quantum treatment the classical observables have to be promoted to the corresponding quantum operators. The quantum Hamiltonian is then
\begin{gather}
H=-{\hbar^2 \over 2m} \nabla^2- {k \over m \rho}\left(-i \hbar {\partial \over \partial z} \right)+{k^2 \over 2m\rho^2}, \label{original}
\end{gather}
where $\nabla^2$ is the Laplacian operator.

It can be seen that the Hamiltonian commutes with the operators of partial derivative with respect to $z$ and $\phi$,
\begin{gather*}
\left[ H, {\partial \over \partial z} \right] = \left[ H, {\partial \over \partial \phi} \right]=0.
\end{gather*}
This suggests us the following ansatz for the solutions of the stationary Schr\"odinger equation
\begin{gather}
\psi(\rho,\phi,z)= e^{i(p_z z+\ell \phi)/ \hbar} \rho^{-1/2}G(\rho), \label{ansatzS}
\end{gather}
where $p_z$ and $\ell$ are respectively the eigenvalues of the momentum operator along $z$, $P_z =-i \hbar   \partial/\partial z$, and the $z$ component of the angular momentum, $L_z =-i \hbar   \partial/\partial\phi$.

Through this ansatz we can separate variables for the stationary Schr\"odinger equation, $H\psi = E \psi$, leading us to the dif\/ferential equation for $G(\rho)$
\begin{gather*}
\left[-{1 \over 2} {\mathrm{d}^2 \over \mathrm{d}\rho^2}+{(\lambda/\hbar)^2-1/4 \over 2 \rho^2}-{p_zk \over \hbar^2 \rho}  \right]G(\rho)={m \over \hbar^2}\left( E - {p_z^2 \over 2m } \right) G(\rho),
\end{gather*}
where $\lambda^2=\ell^2+k^2$. To simplify notation we can express the previous equation as
\begin{gather}
\left[- {1 \over 2}{\mathrm{d}^2 \over \mathrm{d}\rho^2}+{ a(a+1) \over 2 \rho^2}-{b \over \rho}\right] G(\rho)=d G(\rho), \label{schrodinger}
\end{gather}
with
\begin{gather*}
a(a+1)={\lambda^2 \over \hbar^2}-{1 \over 4}, \qquad b= {p_zk \over  \hbar^2}, \qquad d= {mE \over \hbar^2}-{p_z^2 \over 2 \hbar^2 }.
\end{gather*}

\begin{figure}[t]
\centering
\includegraphics[scale=0.84]{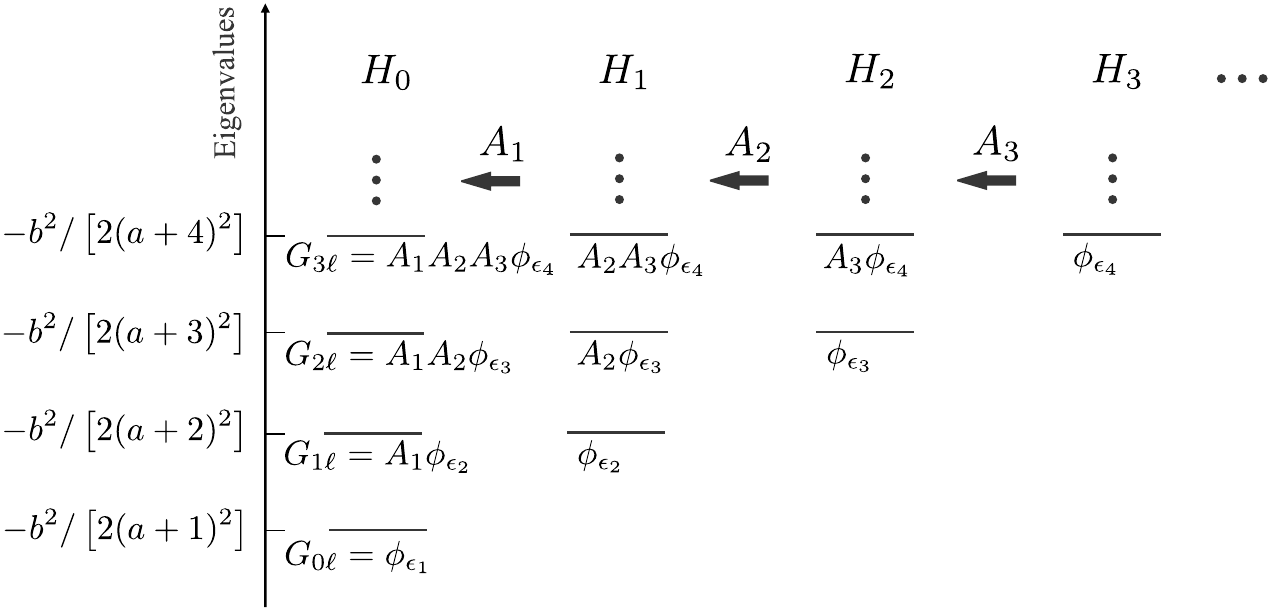}
\caption{A hierarchy of Hamiltonians built up departing from~$H_0$. If we know the ground state of each Hamiltonian and the intertwining operators, we can know the bound states of all Hamiltonians. Note that the eigenvalues are not indeed equidistant. }
\label{fig:Fig1}
\end{figure}

In this work we restrict ourselves to the case $p_z k > 0$, which is the one with bound states. Equation~(\ref{schrodinger}) can be identif\/ied as the radial equation of the hydrogen atom. To solve this equation we propose the existence of a family of operators $A^\dag_n$ that intertwine the Hamiltonians~$H_n$ and~$H_{n+1}$ in the way
\begin{gather*}
H_{n+1} A^\dag_{n+1} = A^\dag_{n+1} H_{n}, 
\end{gather*}
where
\begin{gather*}
H_n   =  - {1 \over 2}{\mathrm{d}^2 \over \mathrm{d}\rho^2}+V_n(\rho) =- {1 \over 2}{\mathrm{d}^2 \over \mathrm{d}\rho^2}+{ (a+n)(a+n+1) \over 2 \rho^2}-{b \over \rho}
\end{gather*}
and
\begin{gather*}
A^\dag_n={1\over \sqrt{2}}\left(-{\mathrm{d} \over \mathrm{d}\rho}+W_n(\rho)\right).
\end{gather*}
From equations~(\ref{ricati}) we have
\begin{gather*}
W_n(\rho)= {a+n \over \rho}-{b \over a+n}, \qquad \epsilon_n = - {b^2 \over 2(a+n)^2}.
\end{gather*}

\begin{figure}[t]
\centering
\includegraphics[scale=0.84]{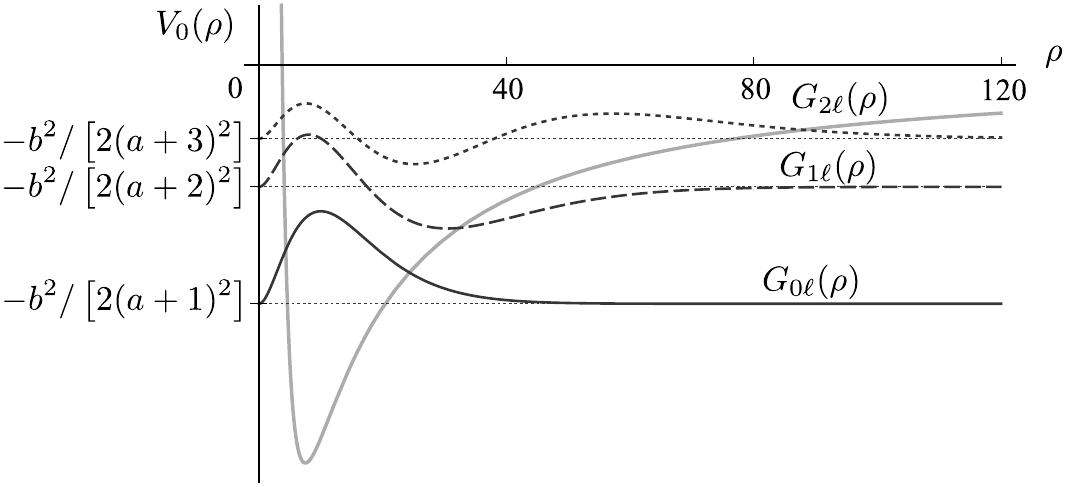}
\caption{The potential $V_0(\rho)$ (gray curve) and its f\/irst three eigenfunctions, $G_{0 \ell}$ (black continuous line), $G_{1 \ell}$ (dashed line) and $G_{2 \ell}$ (dotted line), with $a=1.5$ and $b=0.5$ in units of $1/\rho$. }
\label{fig:Fig2}
\end{figure}

Looking for the function annihilated by $A^\dag_n$, which due to equation~(\ref{factorizacion}) is an eigenfunction of~$H_{n-1}$ with eigenvalue $\epsilon_n$,
\begin{gather*}
A^\dag_n\phi_{\epsilon_{n}}=0 \quad \Rightarrow \quad \phi_{\epsilon_{n}} =  \rho^{a+n}e^{-{b \over a+n}\rho}, 
\end{gather*}
it turns out that the ground state of $H_n$ is given by
\begin{gather*}
\phi_{\epsilon_{n+1}}(\rho) = \rho^{a+n+1}e^{-{b \over a+n+1}\rho}.
\end{gather*}
As expected, one can verify that
\begin{gather}
W_{n+1}(\rho)= \phi'_{\epsilon_{n+1}} (\rho) / \phi_{\epsilon_{n+1}}(\rho). \label{superpotencial}
\end{gather}
It is enough to know the ground state and its eigenvalue for any Hamiltonian of the hierarchy in order to f\/ind the complete solution of $H_0$ (see Fig.~\ref{fig:Fig1}). The spectrum is given by
\begin{gather*}
\mathrm{Sp} \left[H_0 \right] = \left\{ -{b^2 \over 2(a+n+1)^2}, \  n=0,1,2,\dots\right\},
\end{gather*}
and its eigenfunctions by
\begin{gather*}
G_{0 \ell} = \phi_{\epsilon_1}, \qquad G_{1 \ell}= A_1\phi_{\epsilon_2}, \qquad G_{2 \ell}= A_1 A_2 \phi_{\epsilon_3}, \qquad G_{3 \ell}= A_1 A_2 A_3 \phi_{\epsilon_4}, \qquad \dots,
\end{gather*}
where the index $n$ indicates the energy level of $H_0$ and the index $\ell$ reminds us that the radial Hamiltonian depends on the angular momentum. In Fig.~\ref{fig:Fig2} the f\/irst three eigenfunctions of $H_0$ can be seen (black continuous, dashed and dotted lines) placed at its corresponding energy level and the potential~$V_0(\rho)$ is as well drawn (gray line).

Returning to the original problem, i.e.\ the eigenvalue equation for the operator of equation~(\ref{original}), we have that the spectrum is given by
\begin{gather*}
\mathrm{Sp}\left[H \right]=\left\{ {p_z^2 \over 2m} \left[1- {k^2\over \hbar^2 (\lambda/\hbar + n + 1/2)^2} \right], \  n=0,1,2, \dots\right\},
\end{gather*}
and its eigenfunctions by
\begin{gather*}
\psi_{n \ell p_z}(\rho, \phi, z)= C_{n \ell p_z}e^{i(p_z z + \ell \phi )/\hbar  } \rho^{-1/2}G_{n \ell}(\rho),
\end{gather*}
where $C_{n\ell p_z}$ is a normalization constant (there is not sum convention).

\section{Relativistic quantum approach}\label{sec:RQM}

The stationary Dirac equation of a free particle with mass $m$ and spin $1/2$ is
\begin{gather}
 {\bf H}_D \Psi = \big[ c \vec{\alpha} \cdot \vec{{\bf P}}+\beta m c^2  \big] \Psi= E \Psi, \label{Dirac}
\end{gather}
where $\vec{{\bf P}}$ is the  momentum operator, $\Psi$ is a four-component spinor and $\alpha_i$ and $\beta$ are $4 \times 4$ matrices given by
\begin{gather*}
\alpha_i=   \begin{pmatrix}
0 & \sigma_i \\
\sigma_i & 0
\end{pmatrix}, \qquad \beta =   \begin{pmatrix}
\sigma_0 & 0 \\
0 & - \sigma_0
\end{pmatrix},
\end{gather*}
being $\sigma_0$  the $2\times 2$ identity matrix and $\sigma_i$  the  Pauli matrices. The $4 \times 4$ matrix operators are written in boldface  in order to be distinguished  from the $2 \times 2$ matrix operators. In our case the interaction with the magnetic f\/ield
derived from the vector potential $\vec{{\bf A}} = { c k \over e \rho} \hat{e}_z$
is described by  the minimal coupling rule
$\vec{{\bf P}} \rightarrow \vec{{\bf P}}- {e \over c }\vec{{\bf A}}$.
In cylindrical coordinates the resulting stationary Dirac equation is
\begin{gather}\label{heff}
{\bf H}_D \Psi= \left\{ -i \hbar c {\bf D}(\phi)\alpha_1 {\partial \over \partial \rho}- {i \hbar c \over \rho} {\bf D}(\phi) \alpha_2 {\partial \over \partial \phi}- i \hbar c \alpha_3 {\partial \over \partial z} - {k \over \rho} \alpha_3 + \beta m c^2\right\} \Psi = E \Psi,
\end{gather}
where  ${\bf D}(\phi)= \mathrm{Diag} \left [e^{-i\phi},e^{i\phi},e^{-i\phi},e^{i\phi} \right]$ is
a diagonal matrix. This interacting Hamiltonian commutes with the   momentum operator, and with the total angular momentum in the $z$-direction,
\begin{gather*}
{\bf P}_z=-i \hbar \partial_z  \mathbb{1},\qquad
{\bf J}_z=-i \hbar \partial_\phi  \mathbb{1} + {\hbar \over 2} \Sigma_3,\qquad
\Sigma_i=   \begin{pmatrix}
\sigma_i & 0 \\
0 & \sigma_i
\end{pmatrix}.
\end{gather*}
Then we will look for a solution to equation (\ref{heff}) that is also an eigenfunction of these two operators with corresponding
eigenvalues $p_z$ and $\ell$ respectively (see equation~(\ref{ansatzS})), having the form
\begin{gather*}
\Psi(\rho, \phi,z)=e^{i p_z z / \hbar}e^{i(\ell \mathbb{1} -\Sigma_3/2)\phi/\hbar} \rho^{-1/2}G_D(\rho).
\end{gather*}
The equation that the  radial function $G_D(\rho)$ must fulf\/ill is
\begin{gather*}
\left[-i \alpha_1 {\mathrm{d} \over \mathrm{d\rho}}+{\ell \over \hbar \rho}\alpha_2- \left( {k \over \hbar \rho}  - {p_z \over \hbar} \right) \alpha_3 +{mc\over \hbar}\beta \right] G_D(\rho)={E\over c \hbar } G_D(\rho).
\end{gather*}
The operator between brackets  is an ef\/fective Hamiltonian that will be called $\bf{H}_\rho$.
It is useful to perform an unitary transformation  in order to leave all the dependence of $\rho$ on a single matrix,
\begin{gather}
{\bf H}_0={\bf U}_1^\dag {\bf H}_\rho {\bf U}_1, \qquad {\bf U}_1=e^{-i\theta \Sigma_1 /2}= \cos(\theta/2)\mathbb{1} -i\sin(\theta/2)\Sigma_1, \label{rotacion}
\end{gather}
with $\tan \theta = -k/\ell$. In the rotated frame the Hamiltonian is
\begin{gather*}
{\bf H}_0=-i  \alpha_1  {\mathrm{d} \over \mathrm{d} \rho}+ \left( { \lambda \over \hbar \rho} -{ p_z k \over \hbar \lambda}\right) \alpha_2+ { p_z \ell \over \hbar \lambda} \alpha_3+{m c\over \hbar}\beta,
\end{gather*}
where in this context once again $\lambda^2=\ell^2+k^2$.
This Hamiltonian has a special  structure  that can be better appreciated if we write it as follows
\begin{gather*}
{\bf H}_0=  \begin{pmatrix}
(m c /\hbar) \sigma_0  & h_0 \\
h_0 & -(m c /\hbar) \sigma_0
\end{pmatrix}, \qquad h_0= - i\sigma_1  {\mathrm{d} \over \mathrm{d} \rho}+ \left( { \lambda \over \hbar \rho} -{ p_z k \over \hbar \lambda}\right) \sigma_2+ {p_z \ell \over \hbar \lambda} \sigma_3, 
\end{gather*}
being $h_0$ a $2 \times 2$ matrix operator. In order to simplify  notation we will write
\begin{gather*}
h_0=- i\sigma_1  {\mathrm{d} \over \mathrm{d} \rho}+ \left( { a\over  \rho} -{ b_0 \over a}\right) \sigma_2+ d_0 \sigma_3 = - i\sigma_1  {\mathrm{d} \over \mathrm{d} \rho} + v_0(a,b_0,d_0;\rho),
\end{gather*}
with
\begin{gather*}
a=\lambda/\hbar, \qquad b_0=p_z k/\hbar^2,\qquad d_0=p_z \ell/ \hbar \lambda.
\end{gather*}
To solve the eigenvalue equation for ${\bf H}_0 $ with a method similar to that used in the nonrelativistic approach,  we propose the intertwining relationship
\begin{gather}\label{int}
{\bf H}_{n+1} {\bf A}^\dag_{n+1} = {\bf A}^\dag_{n+1} {\bf H}_n,
\end{gather}
with the intertwining operators ${\bf A}^\dag_{n+1}$ and the sequence of Hamiltonians ${\bf H}_{n+1}$
having the form
\begin{gather}
{\bf A}_{n+1}^\dag  =    \begin{pmatrix}
B_{n+1}^\dag  & 0 \\
0 & B_{n+1}^\dag
\end{pmatrix},
\qquad {\bf H}_n =   \begin{pmatrix}
(m c /\hbar)  \sigma_0  & h_n \\
h_n & - (m c /\hbar) \sigma_0
\end{pmatrix} , \label{intertwining}
\end{gather}
where $h_n(a_n,b_n,d_n;\rho)=h_0(a+n,b_n,d_n;\rho)$ with parameters $b_n$ and $d_n$
to be determined and~$B_{n+1}^\dag$ is a $2 \times 2$ operator that intertwines $h_{n+1}$ with $h_n$.
The structure of~${\bf A}_{n+1}^\dag$ above proposed is the simplest choice. A more general form of the intertwining operators  will be presented  elsewhere. Similar $2\times 2$ intertwining operators
were considered in~\cite{ioffe}.

In the same way as in the nonrelativistic quantum case (see equation~(\ref{entrelazamiento})), the f\/irst order  intertwining operator has the following structure
\begin{gather*}
{\bf A}^\dag_{n+1} = - \mathbb{1}  {\mathrm{d} \over \mathrm{d} \rho}+ {\bf W}^\dag_{n+1}(\rho),
\end{gather*}
where ${\bf W}^\dag_{n+1}(\rho)$ is a variable matrix. In the same way $B_{n+1}^\dag$ reads
\begin{gather*}
B_{n+1}^\dag=-\sigma_0 {\mathrm{d} \over \mathrm{d} \rho}+F_{n+1}(\rho),
\end{gather*}
with $F_{n+1}(\rho)$ a $2\times2$ matrix.

Solving the intertwining relation of equation~(\ref{int}) we f\/ind
\begin{gather*}
 b_n=b_0=b,\qquad d_n^2=d_0^2+{n(2a+n)b^2 \over a^2(a+n)^2},\nonumber \\
 B_{n+1}^\dag=-\sigma_0 {\mathrm{d}\over \mathrm{d\rho}}+{2(a+n)+1\over 2}\left[ {1\over \rho} -{b\over (a+n)(a+n+1)} \right] \sigma_0 - {d_{n+1} - d_n \over 2} \left(i \sigma_1- \sigma_2 \right)  \nonumber \\
\hphantom{B_{n+1}^\dag=}{}
- {1\over 2}\left[{1\over \rho}+{b \over (a+n)(a+n+1)} \right] \sigma_3.  
\end{gather*}
Then the intertwining operators ${\bf A}^\dag_{n+1}$ are directly obtained by substitution in equation~(\ref{intertwining}).

Now, we look for the vector
wavefunctions annihilated by ${\bf A}_{n+1}^\dag$. Taking advantage of the block diagonal structure of this operator,  f\/irst we f\/ind  the two-component functions annihilated by $B^\dag_{n+1}$.
These are given by the following two independent vector functions
\begin{gather*}
\chi_{n} =    \begin{pmatrix}
1 \\
0
\end{pmatrix} \rho^{a+n} e^{-b \rho/(a+n)}, 
\\
\xi_n =    \begin{pmatrix}
i \dfrac{(a+n)^2(a+n+1)^2(d_{n+1}-d_n)}{b^2} \left(1-\dfrac{b}{(a+n)(a+n+1)}\rho\right)  \vspace{1mm}\\
\rho
\end{pmatrix} \rho^{a+n} e^{-b \rho/(a+n+1)}.
\end{gather*}
We can built four-component functions annihilated
by ${\bf A}_{n+1}^\dag$ such that at the same time they are also eigenvectors of ${\bf H}_n$.
We f\/ind four types of such vectors and the corresponding eigenvalues:
\begin{alignat*}{3}
& \phi_{an} =   \begin{pmatrix}
\chi_n \vspace{1mm}\\
\dfrac{d_n}{\sqrt{(m c /\hbar)^2+d_n^2}+m c /\hbar} \chi_n
\end{pmatrix},
\qquad & &\sqrt{(m c /\hbar)^2+d_n^2}; &
\\
& \phi_{bn}=   \begin{pmatrix}
\chi_n \vspace{1mm}\\
-\dfrac{d_n}{\sqrt{(m c /\hbar)^2+d_n^2}-m c /\hbar } \chi_n
\end{pmatrix},
\qquad & &-\sqrt{(m c /\hbar)^2+d_n^2}; &
\\
& \phi_{cn}=   \begin{pmatrix}
\xi_n \vspace{1mm}\\
-\dfrac{d_{n+1}}{\sqrt{(m c /\hbar)^2+d_{n+1}^2}+m c /\hbar} \xi_n
\end{pmatrix},
\qquad & &\sqrt{(m c /\hbar)^2+d_{n+1}^2}; &
\\
& \phi_{dn}=  \begin{pmatrix}
\xi_n \vspace{1mm}\\
\dfrac{d_{n+1}}{\sqrt{(m c /\hbar)^2+d_{n+1}^2}-m c /\hbar} \xi_n
\end{pmatrix},
\qquad  &&-\sqrt{(m c /\hbar)^2+d_{n+1}^2}. &
\end{alignat*}
We will brief\/ly comment here on some properties of these results,  a more complete discussion will be given elsewhere.
\begin{itemize}\itemsep=0pt
\item
{\em Superpotential}.
Consider now the matrix $\Xi_{n+1}(\rho)$ which is constructed by placing the vectors $\phi_{\nu n}$ in its columns, it can be verif\/ied, in analogy with equation~(\ref{superpotencial}), that
\begin{gather*}
{\bf W}^\dag_{n+1}(\rho)=-\Xi_{n+1}'(\rho) \Xi_{n+1}^{-1}(\rho).
\end{gather*}

\item
{\em Spectrum}.
From the intertwining relationship, equation~(\ref{int}), it is shown  that the spectrum of ${\bf H}_0$ is given by
\begin{gather*}
\mathrm{Sp} \left[{\bf H}_0  \right] = \pm \sqrt{\left({m c \over \hbar}\right)^2+d_0^2+{n(2a+n)b^2\over a^2(a+n)^2}},\qquad n=0,1,\dots. 
\end{gather*}

\item
{\em Eigenfunctions}. The eigenfunctions of the
initial Hamiltonian are computed in the usual way. We have four types of eigenfunctions:
\begin{alignat*}{5} 
&  \Phi_{a0\ell}= \phi_{a0}, \qquad &&\Phi_{a1\ell}= {\bf A}_1 \phi_{a1}, \qquad &&\Phi_{a2\ell}= {\bf A}_1 {\bf A}_2 \phi_{a2}, \qquad && \dots , & \\
&  \Phi_{b0\ell}= \phi_{b0}, \qquad &&\Phi_{b1\ell}= {\bf A}_1 \phi_{b1}, \qquad &&\Phi_{b2\ell}= {\bf A}_1 {\bf A}_2 \phi_{b2}, \qquad && \dots , & \\
&  \Phi_{c0\ell}= \phi_{c0}, \qquad &&\Phi_{c1\ell}= {\bf A}_1 \phi_{c1}, \qquad &&\Phi_{c2\ell}= {\bf A}_1 {\bf A}_2 \phi_{c2}, \qquad && \dots, & \\
&  \Phi_{d0\ell}= \phi_{d0}, \qquad &&\Phi_{d1\ell}= {\bf A}_1 \phi_{d1}, \qquad &&\Phi_{d2\ell}= {\bf A}_1 {\bf A}_2 \phi_{d2}, \qquad && \dots, & 
\end{alignat*}
where we add the subindex $\ell$ to remind that the Hamiltonian ${\bf H}_0$ depends on $\ell$.
\end{itemize}

\begin{figure}[t]
\centering
\includegraphics[scale=0.84]{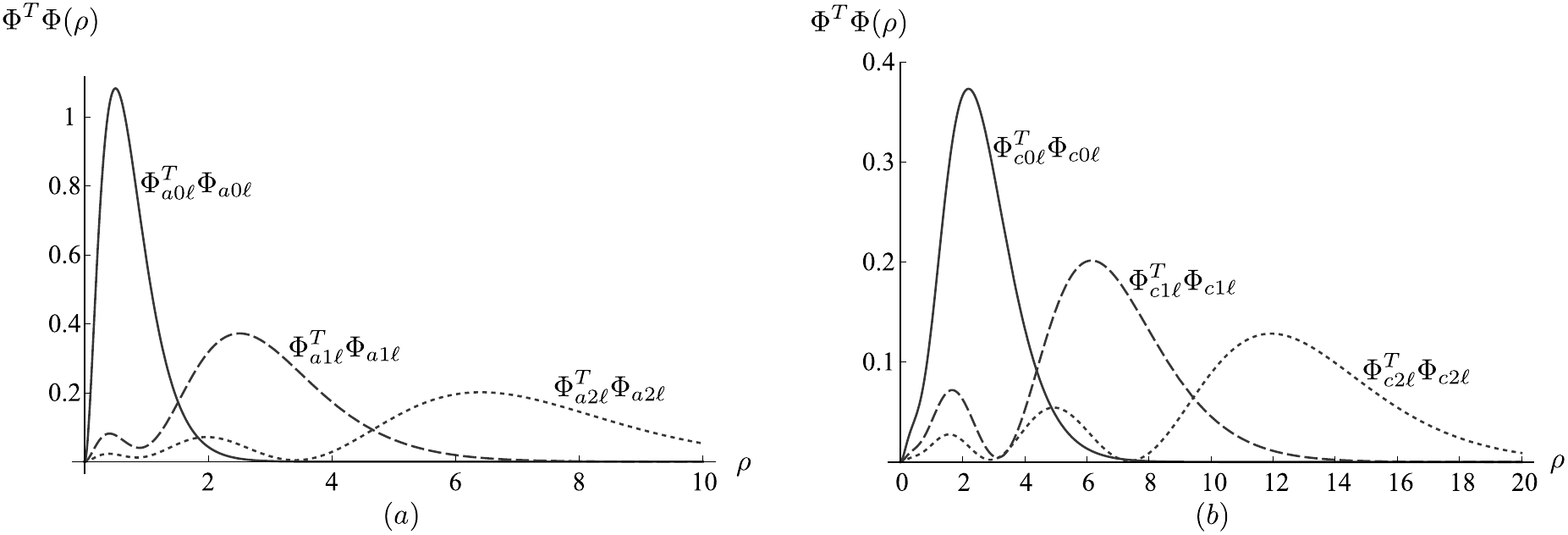}
\caption{Probability densities for six eigenvectors of~${\bf H}_0$: in $(a)$ the f\/irst three of the family $\Phi_{an\ell}$, and in $(b)$ the f\/irst three of the family~$\Phi_{cn\ell}$, the parameters used were $a=1$; $b=2$ and $d_0=1$ with units of~$1/\rho$; $m=0.1$ and $c=\hbar=1$.}
\label{fig:Fig3}
\end{figure}

Then the eigenvectors of the Dirac equation, equation~(\ref{Dirac}), for our system are
\begin{gather*}
\Psi_{\nu n \ell p_z}(\rho,\phi,z)= C_{\nu n \ell p_z} e^{ip_z z/\hbar}e^{i(\ell \mathbb{1}-{1 \over 2}\Sigma_3)\phi / \hbar}{\bf U}_1 \rho^{-1/2} \Phi_{\nu n \ell}(\rho),
\end{gather*}
where ${\bf U}_1$ is given by equation~(\ref{rotacion}), $\nu=a,b,c,d$, $n=0,1,2,\dots$ and $C_{\nu n \ell p_z}$ are normalization constants (there is not sum convection).
The spectrum of ${\bf H}_D $ is $c \hbar$ times the one of ${\bf H}_0$, thus
\begin{gather*}
\mathrm{Sp} \left[{\bf H}_D  \right] = \pm m c^2 \sqrt{1+{p_z^2 \over m^2 c^2}-{p_z^2 k^2 \over \hbar^2 m^2 c^2 (\lambda/\hbar + n)^2}}, \qquad n=0,1,2, \dots. 
\end{gather*}
In Fig.~\ref{fig:Fig3} \looseness=-1
 we can see the probability densities of six eigenvectors of ${\bf H}_0$. In $(a)$ we have the f\/irst three with subindex~$a$, and in~$(b)$ the corresponding but with subindex $c$. Note that $\Phi_{a1\ell}$ and~$\Phi_{c0\ell}$ have the same eigenvalue but the behavior of the probability density is quite dif\/ferent, the same happens with $\Phi_{a2\ell}$ and~$\Phi_{c1\ell}$, and so on. This degeneracy, which does not appear in the nonrela\-tivistic approach, is due to the spin degree of freedom and will be analyzed in detail elsewhere.

\section{Conclusions}\label{C:conclusiones}

In this work we have adapted the intertwining technique to solve exactly the Dirac equation  associated to a charged particle of spin $1/2$ immersed in a magnetic f\/ield with cylindrical symmetry generated by the vector potential $\vec{{A}} = { c k \over e \rho} \hat{e}_z$. We f\/irst addressed the problem in the nonrelativistic regime, i.e., the Schr\"odinger equation through the standard intertwining technique. Afterwards we set up the corresponding Dirac equation and we proposed, as in the nonrelativistic approach, a hierarchy of shape invariant Hamiltonians intertwined by some operators to be determined. These operators afterwards were found and using them the ground states of each Hamiltonian were built. Applying these operators onto the ground states all the bound states were obtained as well as their respective eigenvalues of the original Dirac equation. As far as we know these solutions have not been reported before.  The analogies between the method for solving the Dirac equation and the standard intertwining technique for the Schr\"odinger equation were recurrently employed throughout the entire procedure.

\subsection*{Acknowledgments}
We acknowledge f\/inancial support from Ministerio de Ciencia
e Innovaci\'o—n (MICINN) of Spain, projects MTM2009-10751,
and FIS2009-09002. ACA acknowledges to Conacyt a PhD grant and the kind hospitality at  University of Valladolid. DJFC acknowledges the f\/inancial support of Conacyt, project 152574.

\pdfbookmark[1]{References}{ref}
\LastPageEnding

\end{document}